\documentclass[useAMS,usenatbib]{mn2e}
\usepackage{aas_macros}
\usepackage{graphicx}
\usepackage{amssymb}
\usepackage{amsmath}
\usepackage{color}
\newcommand{\lya}{Lyman-$\alpha$~}

\newcommand{\kms} {km $\rm{s^{-1}}$}

\newcommand{\msolar} {$\rm{M_{\odot}}~$}

\newcommand{\molH} {$\rm{H_2}$~}
\newcommand{\molHc} {$\rm{H_2}$}

\newcommand{\JJ} {\rm{$J_{21}$}~}
\newcommand{\JJc} {\rm{$J_{21}$}}

\begin{document}

\title[DCBH formation under streaming velocities]
{The formation of direct collapse black holes under the influence of streaming velocities}
\author[Schauer et al.]{Anna T. P. Schauer$^{1}
$\thanks{E-mail:schauer@uni-heidelberg.de},
John Regan$^{2,1}$, 
Simon C. O. Glover$^{1}$, Ralf S. Klessen$^{1,3}$\\
$^{1}$ Universit\"at Heidelberg, Zentrum f\"ur Astronomie, Institut f\"ur Theoretische
Astrophysik, Albert-Ueberle-Str. 2, 69120 Heidelberg, Germany\\
$^{2}$ Centre for Astrophysics \& Relativity, School of Mathematical Sciences, Dublin City University, Glasnevin, D09 W6Y4, Dublin, Ireland \\
$^{3}$ Universit\"{a}t Heidelberg, Interdisziplin\"{a}res Zentrum f\"{u}r Wissenschaftliches Rechnen, Im Neuenheimer Feld 205, 69120 Heidelberg, Germany\\}

\pagerange{\pageref{firstpage}--\pageref{lastpage}} \pubyear{2002}

\maketitle

\label{firstpage}

\begin{abstract}
We study the influence of a high baryonic streaming velocity on the formation of direct collapse black holes (DCBHs) 
with the help of cosmological simulations carried out using the moving mesh code {\sc arepo}. We show that a streaming
velocity that is as large as three times the root-mean-squared value is effective at suppressing the formation of H$_{2}$-cooled
minihaloes, while still allowing larger atomic cooling haloes (ACHs) to form. We find that enough H$_{2}$ forms in the centre
of these ACHs to effectively cool the gas, demonstrating that a high streaming velocity by itself cannot produce the conditions
required for DCBH formation. However, we argue that high streaming velocity regions do provide an ideal environment for
the formation of DCBHs in close pairs of ACHs (the ``synchronised halo'' model). Due to the absence of star formation in
minihaloes, the gas remains chemically pristine until the ACHs form. If two such haloes form with only a small separation in
time and space, then the one forming stars earlier can provide enough ultraviolet radiation to suppress H$_{2}$ cooling in
the other, allowing it to collapse to form a DCBH. Baryonic streaming may therefore play a crucial role in the formation of
the seeds of the highest redshift quasars.
\end{abstract}


\section{Introduction}
Massive black hole seeds that can form through gravitational collapse of very massive progenitors, 
\citep{chandrasekhar64}, can be invoked to
explain the observations of quasars at very early times in the Universe \citep{wu15, mort11, fan06}.
The direct collapse (DC) model \citep{loeb94, oh02, bl03, begel06, rh09, rh09a}
of super-massive black hole (SMBH) formation requires that massive objects
form in near pristine atomic cooling haloes in which Population III (Pop III) star formation has
been suppressed. The suppression of previous episodes of star formation means that metal enrichment is obviated.
Pop III star formation can be restrained by removing the primary gas coolant, \molHc, or alternatively by increasing
the minimum halo mass required for collapse due to the impact of streaming motions as we shall see.\\
\indent  One avenue that has been explored by many authors to suppress \molH formation 
is radiative feedback from the surrounding galaxies in the Lyman-Werner (LW) band 
\citep{sbh10, agarw12, regan14b, latif14a, agarwal14, agar15, hartwig15a, regan16a}. 
LW radiation readily dissociates \molH through the two step Solomon
process \citep{field66, stecher67}. 
For high star formation efficiencies or single, massive Pop III stars, most 
of the LW radiation can escape the host galaxy or minihalo \citep{anna15,anna17}.
If the intensity of the LW radiation is sufficient, \molH is dissociated and cooling is suppressed
in low mass minihaloes. The halo then continues to accrete mass all the way up to and beyond the
atomic line cooling limit. Halo collapse can then begin once \lya line cooling becomes effective
and contraction proceeds isothermally at approximately $T_{\rm{gas}} \sim 8000$ K. In this case,
the strength of the background required depends both on the spectral shape of the sources
and on their proximity with respect to the target halo
\citep{dijkstra08, sug14, bhaskar15}. There is general consensus within the literature that the
critical value of the background required to suppress \molH formation throughout the entire halo
is $J_{\rm{crit}} \sim 1000 $ \JJ for a background dominated by emission from stars with an
effective temperature  $T_{\rm eff} \sim 5 \times 10^4$ K,
where \JJ $ = 10^{-21}$ erg cm$^{-2}$ s$^{-1}$ Hz$^{-1}$ sr$^{-1}$.
If there is also a non-negligible X-ray background, then the required value can be even higher
\citep{it15,g16, regan16b}. A much milder intensity level, say $J \lesssim 100 $ \JJc, is enough to delay Pop III formation
until a halo reaches the atomic cooling limit, but in this case \molH readily forms in the
self-shielded core of the ACH, resulting in rapid Pop III formation \citep[e.g.][]{fernandez14, regan17}. 

The value of $J_{\rm crit}$ is orders of magnitude higher than the typical strength of the Lyman-Werner background
at the relevant redshifts, even if one accounts for clustering of sources \citep{ahn09}. Radiation fields of the required
strength will therefore only be encountered in unusual circumstances. One promising way in which a field of the required
strength can be produced is the synchronised halo model \citep{dijkstra08,vis14c,regan17,agarwal17}.
This model supposes that occasionally two ACHs will be found in close proximity. If both haloes remain
chemically pristine until the onset of atomic cooling, and if the collapse times of the haloes are sufficiently similar, then Pop III
star formation can begin in the first halo before the second has finished collapsing. The resulting irradiation from the more
evolved system provides a sufficiently strong LW radiation field to suppress cooling in the secondary (proto-)galaxy, which 
therefore becomes an ideal location for the formation of a DCBH. However, for this model to work, it is necessary to suppress
H$_{2}$ formation in both haloes until they become massive enough to atomically cool. If this is accomplished via LW feedback,
the required field strength is $J \gtrsim 100$ \JJc, which is still somewhat higher than the typical strength of the LW background.

Another, perhaps more natural, mechanism for delaying Pop III star formation is via streaming of the baryons with respect to the dark matter.
This effect, elucidated by \cite{th10}, is a result of the initial offset between the baryonic velocites and the dark matter velocities after recombination.
The velocity offset between baryons and dark matter decays as $\Delta v \propto (1 + z)$, but offsets as large as 9 \kms are still possible at $z \sim 100$.
The impact of this streaming motion on Pop III star formation have been investigated by several authors with general agreement that the minimum halo mass
required for Pop III star formation is increased in the
presence of large-scale streaming velocities \citep[][Schauer et al. 2017b in prep]{stacy11a, greif11, naoz13} as the baryons require a larger gravitational potential in which they can virialise due to the 
additional streaming velocity. 
Therefore, streaming motions 
are not thought
to have a large effect on ACHs illuminated with LW radiation fields with strengths greater than $J_{\rm crit}$ (i.e.\ haloes which would form DCBHs 
even without the streaming; see \citealt{lns14} for details). However, \citet{tm14} argue that very large streaming velocities may suppress Pop III star formation 
entirely in some haloes, allowing pristine haloes above the atomic cooling limit to form, thereby providing ideal locations in which DCBHs can form without requiring 
an extremely strong LW radiation field. This model has been criticised by \cite{vis14b}, who concede that streaming can suppress H$_{2}$ formation in minihaloes
but argue that in ACHs, H$_{2}$ formation in dense collapsing gas is inevitable, regardless of the streaming velocity, unless the gas is illuminated by a strong
LW radiation field. Other models which can successfully suppress \molH cooling and delay PopIII formation have been put forward by \cite{io12} (collisional
  dissociation by shocks), \cite{fernandez14} (rapid accretion in a mild LW background) and
\cite{inayoshi15} (proto-galactic collisions) among other. However, here we focus on streaming velocities as a mechanism for increasing the minimum mass for star formation and hence for creating
ideal environments for DCBH formation.

In this paper, we re-examine the role of baryonic streaming in the formation of DCBHs. We show that although baryonic streaming by itself cannot produce haloes capable
of forming DCBHs, it provides a very natural mechanism for suppressing H$_{2}$ formation in pairs of haloes until they reach the atomic cooling regime.
It therefore offers a simple way of producing the pristine ACH pairs required by the synchronised halo model without the need for a locally elevated LW background. 
Furthermore, the risk of metal enrichment from background galaxies is significantly reduced since mini-halo formation is suppressed in the immediate vicinity.

\section{Simulations}
\subsection{Numerical method}
The simulations described in this paper are carried out using the {\sc arepo} moving-mesh code \citep{arepo}. 
We include both dark matter particles and gas cells. The hydrodynamics of the gas is evolved using 
a Voronoi grid in which the gas cells move along with the flow. 
We use a recent version of {\sc arepo} that includes improvements to the time integration scheme, spatial gradient reconstruction and grid regularization discussed in \citet{pakmor16} and \citet{mocz15}. We model the thermal and chemical evolution of the gas in our cosmological boxes using an updated version of the chemical network and cooling function implemented in {\sc arepo} by \citet{hartwig15a}, based on earlier work by \citet{gj07} and \citet{cgkb11}. Our chemical model accounts for the formation of H$_{2}$ by the H$^{-}$ and H$_{2}^{+}$ pathways and its destruction by collisions and by photodissociation. It includes all of the chemical reactions identified by \citet{glover15} as being important for accurately tracking the H$_{2}$ abundance in simulations of DCBH formation. 

Compared to \citet{hartwig15a}, we have made three significant changes to the chemical network. First, we have included a simplified treatment of deuterium chemistry designed 
to track the formation and destruction of HD following \citet{cgkb11}.
Second, we have updated the rate coefficient used for the dielectronic recombination of ionized helium, He$^{+}$; we now use the rate calculated by \citet{b06}. Finally, we have improved our treatment of H$^{-}$ photodetachment to account for the contribution to the H$^{-}$ photodetachment rate coming from non-thermal CMB photons produced as a consequence of cosmological recombination \citep{hp06}. We describe this process using the simple analytical expression given in \citet{cop11}.

Our treatment of the cooling accounts for all of the processes important at the number densities probed by our simulations, including Lyman-$\alpha$ cooling and H$_{2}$ rotational and vibrational cooling. The latter is taken from \citet{ga08} and updated as described in \citet{glover15}.
Further details of our chemical and thermal treatment can be found in Schauer et~al.~(2017, in prep).

\subsection{Initial conditions}
We initialise our simulations at $z = 200$. We assume a $\Lambda$CDM cosmology and use cosmological parameters from the 2015 Planck data release \citep{Planck15}. 
Unless we explicitly state otherwise, we use physical units throughout the paper. 
The initial conditions for the dark matter are created with {\sc music} \citep{hahn11},  using the transfer functions of \cite{eh98}. The baryons are assumed to initially trace the dark matter density distribution. To account for the supersonic streaming of the baryons relative to the dark matter, we first set up a velocity field in the baryonic component that is the same as that in the dark matter, and then impose a constant velocity offset, $v_{\rm stream}$. We take the magnitude of this offset to be $v_{\rm stream} = 18$~km~s$^{-1}$ at $z=200$, corresponding to a value of 3 $\sigma_{\rm rms}$, of the root-mean-squared streaming velocity at this redshift \citep{th10}.

We carry out two simulations, runs SB (small box) and LB (large box). In run SB, we model a box of size ($1 \, {\rm Mpc}/h$) in comoving units, while in run LB, we consider a box of size ($4 \, {\rm Mpc}/h$). Both simulations are performed using 1024$^3$ dark matter particles and initially represent the gas using $1024^{3}$ Voronoi mesh cells. In run SB, we therefore have a dark matter particle mass of $99.4 \, {\rm M_{\odot}}$ and an average initial mesh cell mass of $18.6 \, {\rm M_{\odot}}$. In run LB, these values are a factor of 64 larger, as summarised in Table~\ref{tab:sim}. If we conservatively require 1000 dark matter particles in order to consider a halo to be resolved \citep[][Schauer et al. 2017b in prep.]{mei14}, then the corresponding resolution limits are $M_{\rm DM, res} \simeq 10^{5} \, {\rm M_{\odot}}$ for run SB and $M_{\rm DM, res} \simeq 6.4 \times 10^{6} \, {\rm M_{\odot}}$ for run LB. We note that even in our lower resolution, larger box run, the minimum resolvable halo mass is smaller than the minimum mass of an ACH in the range of redshifts considered in this paper, $M_{\rm atom} \sim 10^{7} \: {\rm M_{\odot}}$. 

\begin{table}
\begin{center}
\begin{tabular}{lccc}
Name  & Box length  & $M_{\rm DM}$ (M$_{\odot}$) & $M_{\rm gas, init}$ (M$_{\odot}$) \\
& ($h^{-1} \, {\rm Mpc}$) & & \\
\hline
SB & 1 & 99 & 18.6 \\
LB & 4 & 6360 & 1190 \\
\hline
\end{tabular}
\caption{Dark matter particle mass and initial mesh cell gas mass in our two simulations.  \label{tab:sim}}
\end{center}
\end{table}

For gas number densities $n < 100 \, {\rm cm^{-3}}$, we adopt a ``constant mass'' refinement criterion, meaning that {\sc arepo} refines or de-refines mesh cells as required in order to keep the mass of gas in each cell at its initial value, plus or minus some small tolerance. At $n \geq 100 \, {\rm cm^{-3}}$, we instead use Jeans refinement, and ensure that the Jeans length is always resolved by at least eight mesh cells. However, in order to prevent run-away collapse to protostellar densities, we switch off refinement once the cell volume becomes less than $0.1 \: h^{-3} \, {\rm pc^{3}}$ in comoving units, corresponding to a density of $n \sim 10^{6} \: {\rm cm^{-3}}$.

\section{Results}
\label{res}
\subsection{Masses and collapse redshifts of haloes containing cold dense gas}
We expect runaway collapse for densities exceeding a number density of  $n \ge 10^4$\,cm$^{-3}$ \citep{glover05}. 
We therefore call a halo collapsed when this number density is exceeded by at least one gas cell in that halo. In our simulations, the first objects collapse at $z = 19$ in 
simulation LB and at $z = 14$ in simulation SB. This difference in collapse redshift is an expected consequence of the difference in box size: simulation LB contains
more large-scale power and hence naturally forms collapsed objects somewhat earlier than simulation SB.

\begin{figure}
\includegraphics[width=0.99\columnwidth]{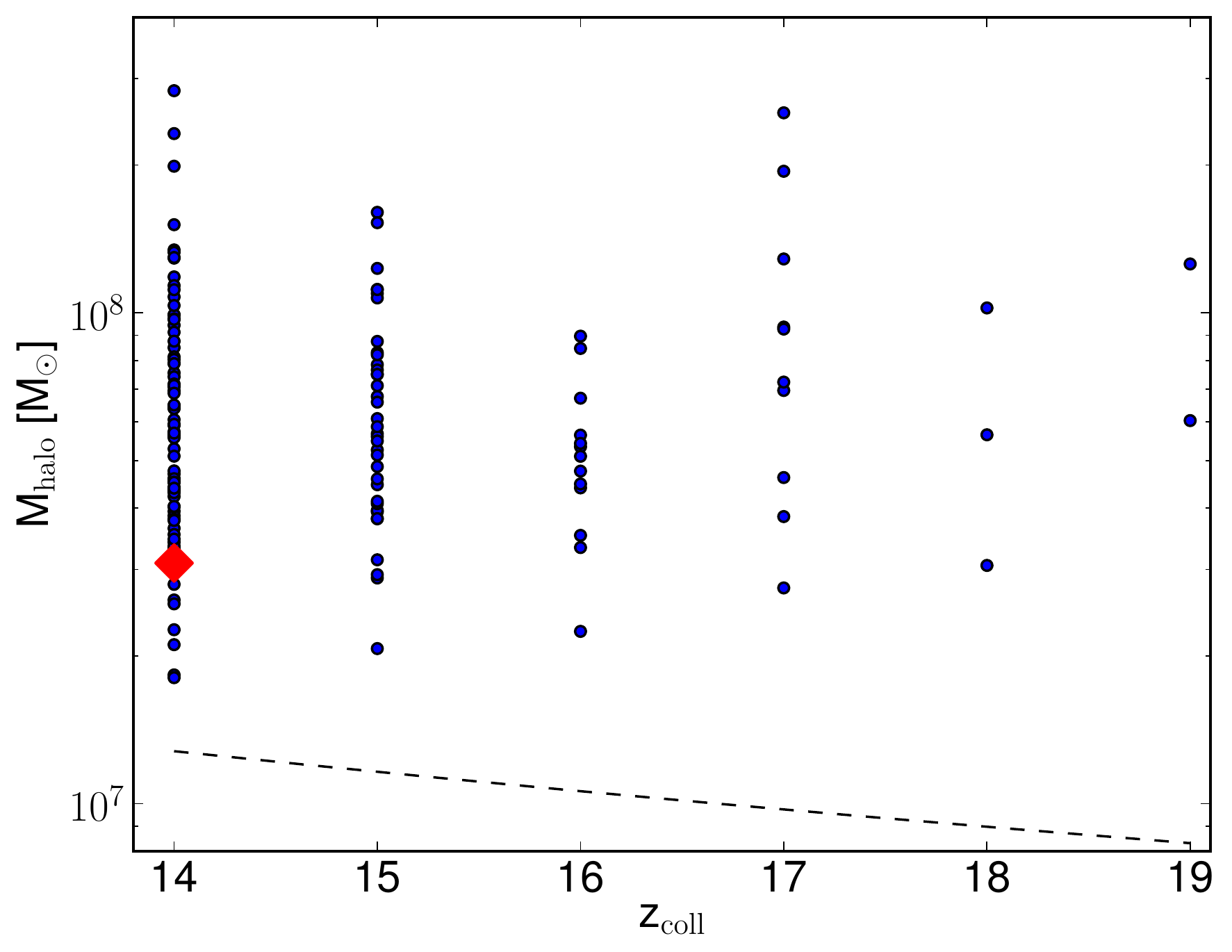}
\caption{Masses of haloes containing cold dense gas, plotted at their redshift of collapse $z_{\rm coll}$, defined as described in Section~\ref{res}. The clustering of points at particular values of $z_{\rm coll}$ reflects the fact that we produce output snapshots with a redshift separation $\Delta z = 1$. The red diamond denotes the halo in simulation SB, and the blue circles denote the haloes in simulation LB.  All collapsed haloes have masses larger than the minimum mass of an ACH (denoted by the dashed black line), demonstrating that only ACHs can collapse when $v_{\rm stream} = 3 \sigma_{\rm rms}$.}
\label{fig:coll}
\end{figure}

In Figure \ref{fig:coll}, we show the halo mass at collapse as a function of collapse redshift $z_\mathrm{coll}$. In our simulations, we produce output snapshots with a redshift spacing $\Delta z = 1$ in order to restrict the volume of data produced to a reasonable level. We can therefore not distinguish between haloes forming at e.g.\ $z = 19.99$ and $z = 19.01$, leading to the clear clustering of points in the Figure. 

\begin{figure*}
\includegraphics[width=0.99\columnwidth]{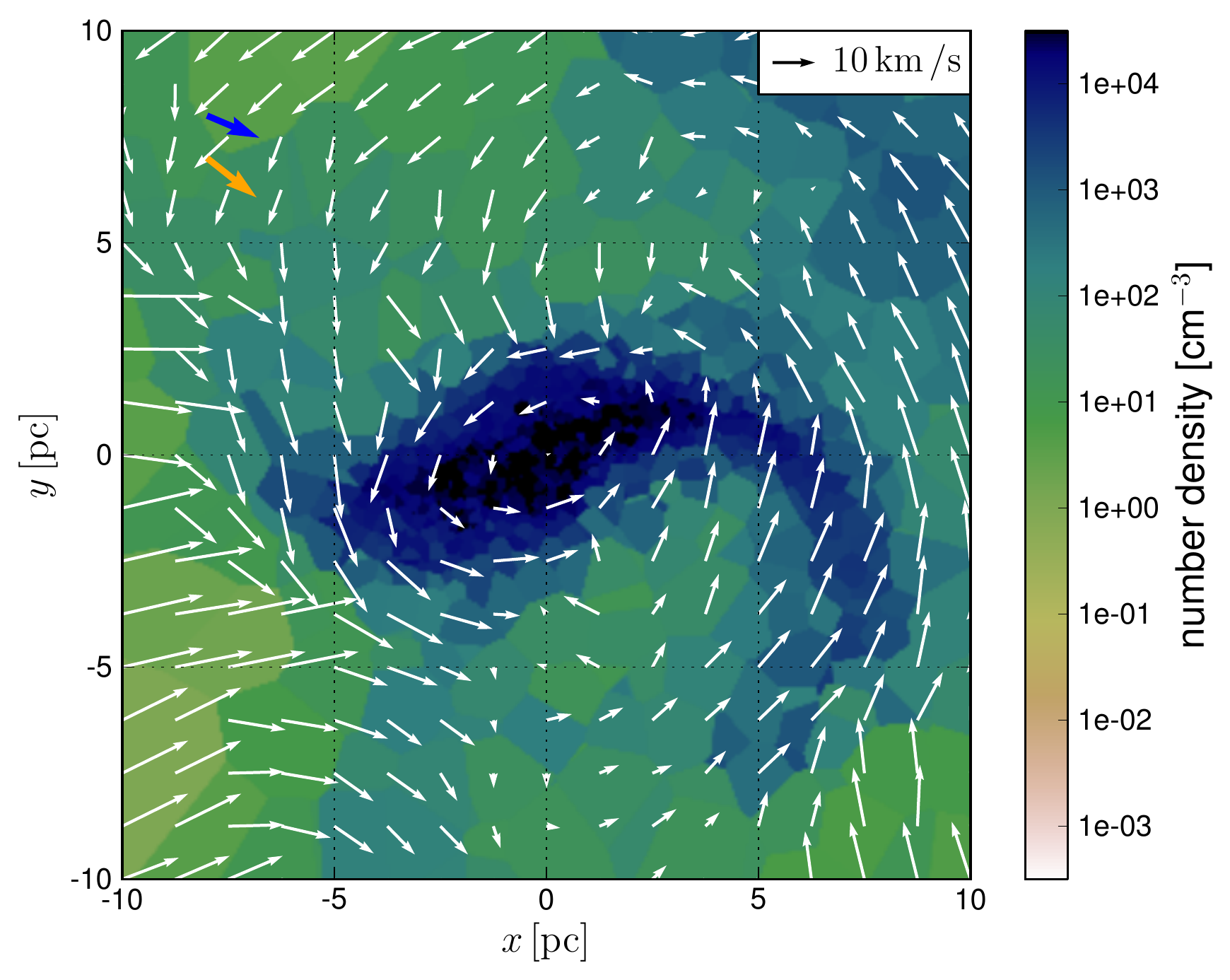}
\includegraphics[width=0.99\columnwidth]{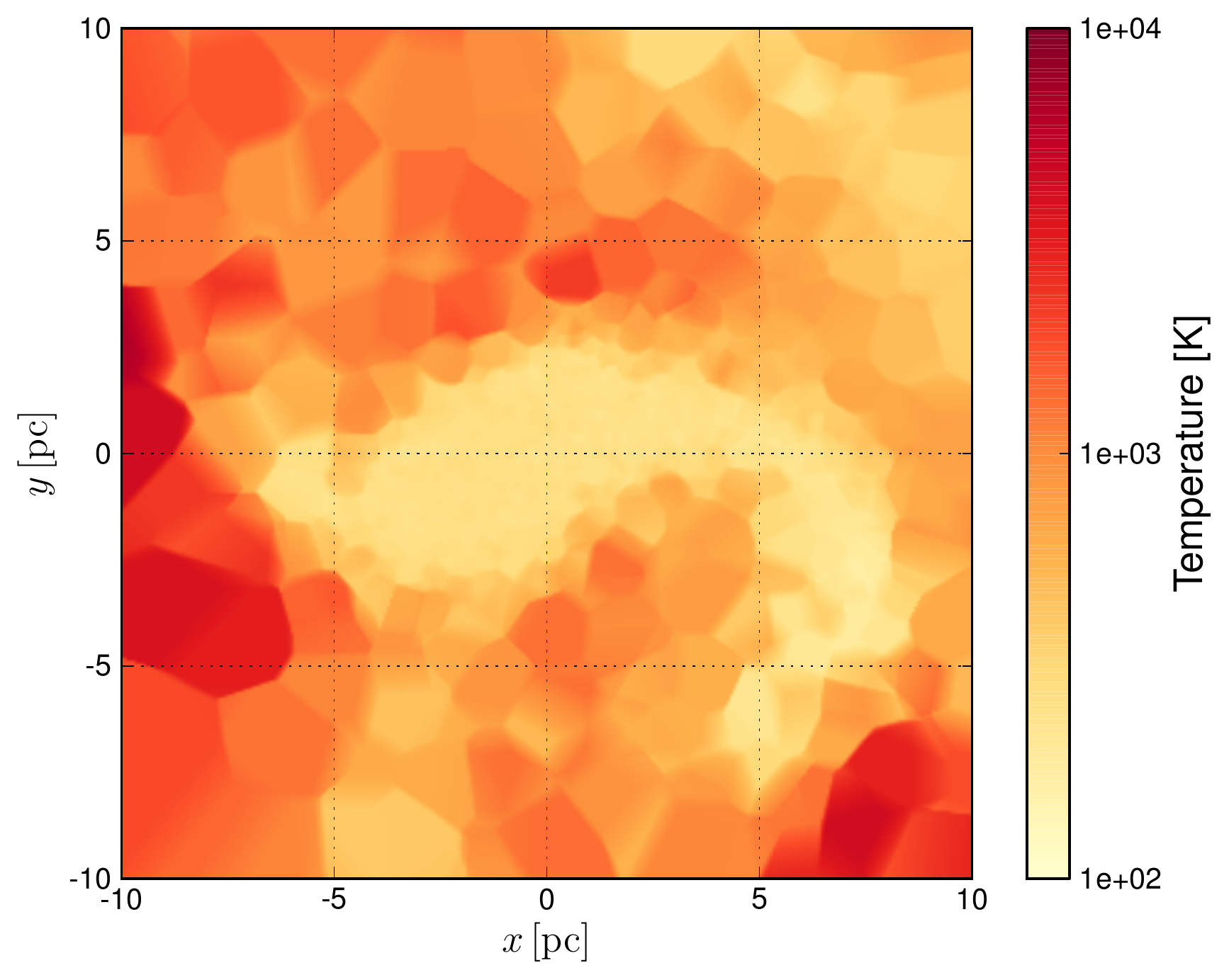}
\includegraphics[width=0.99\columnwidth]{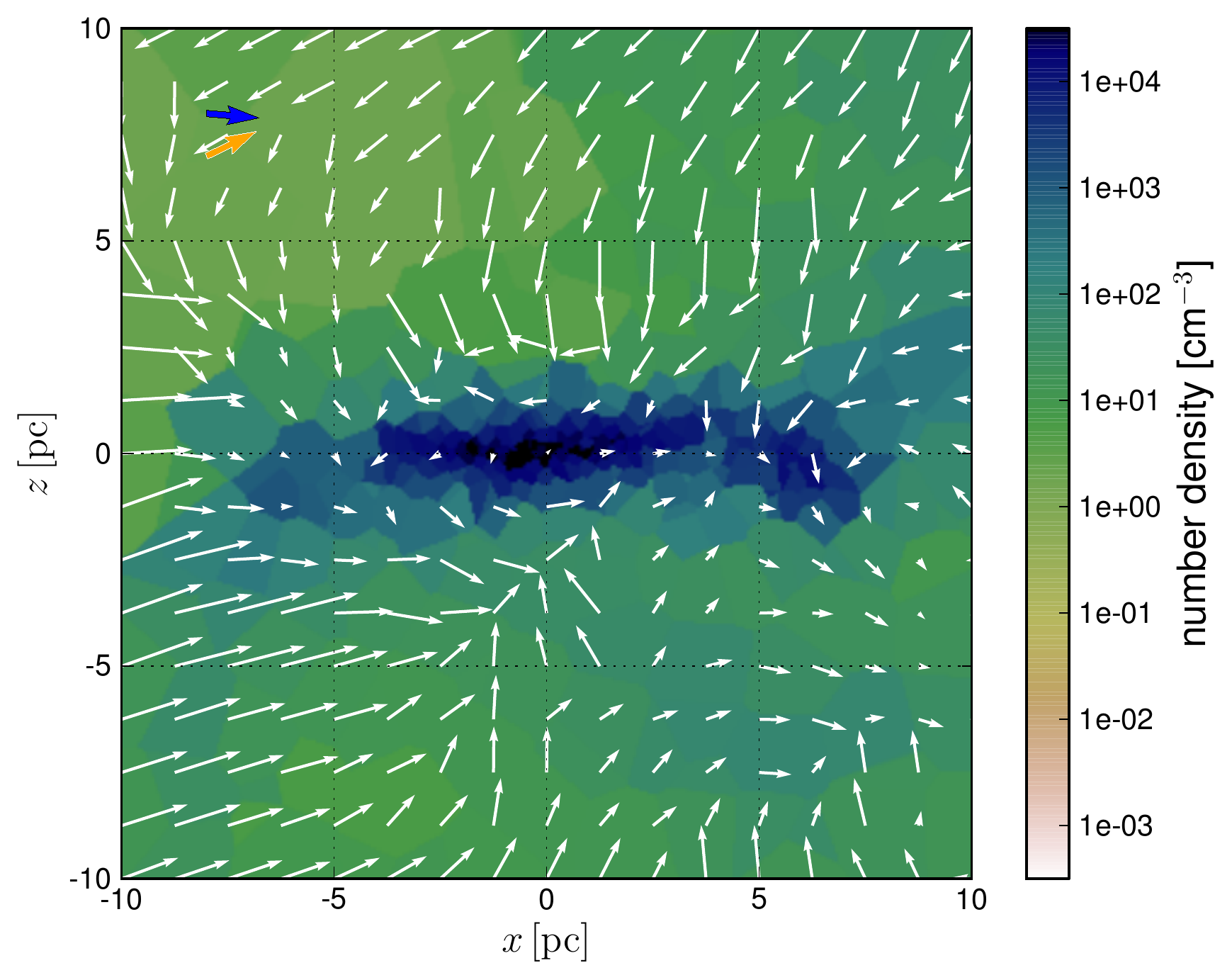}
\includegraphics[width=0.99\columnwidth]{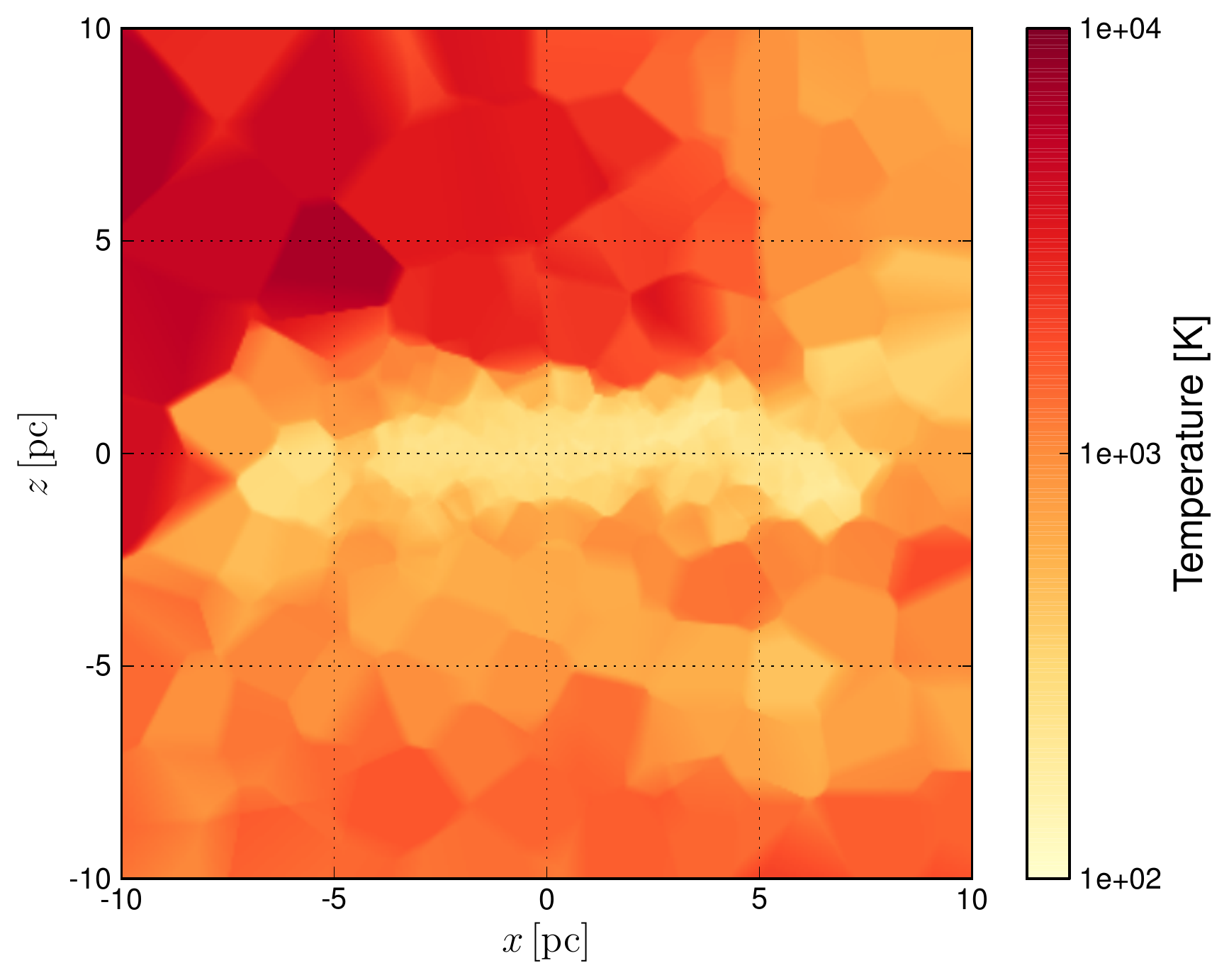}
\includegraphics[width=0.99\columnwidth]{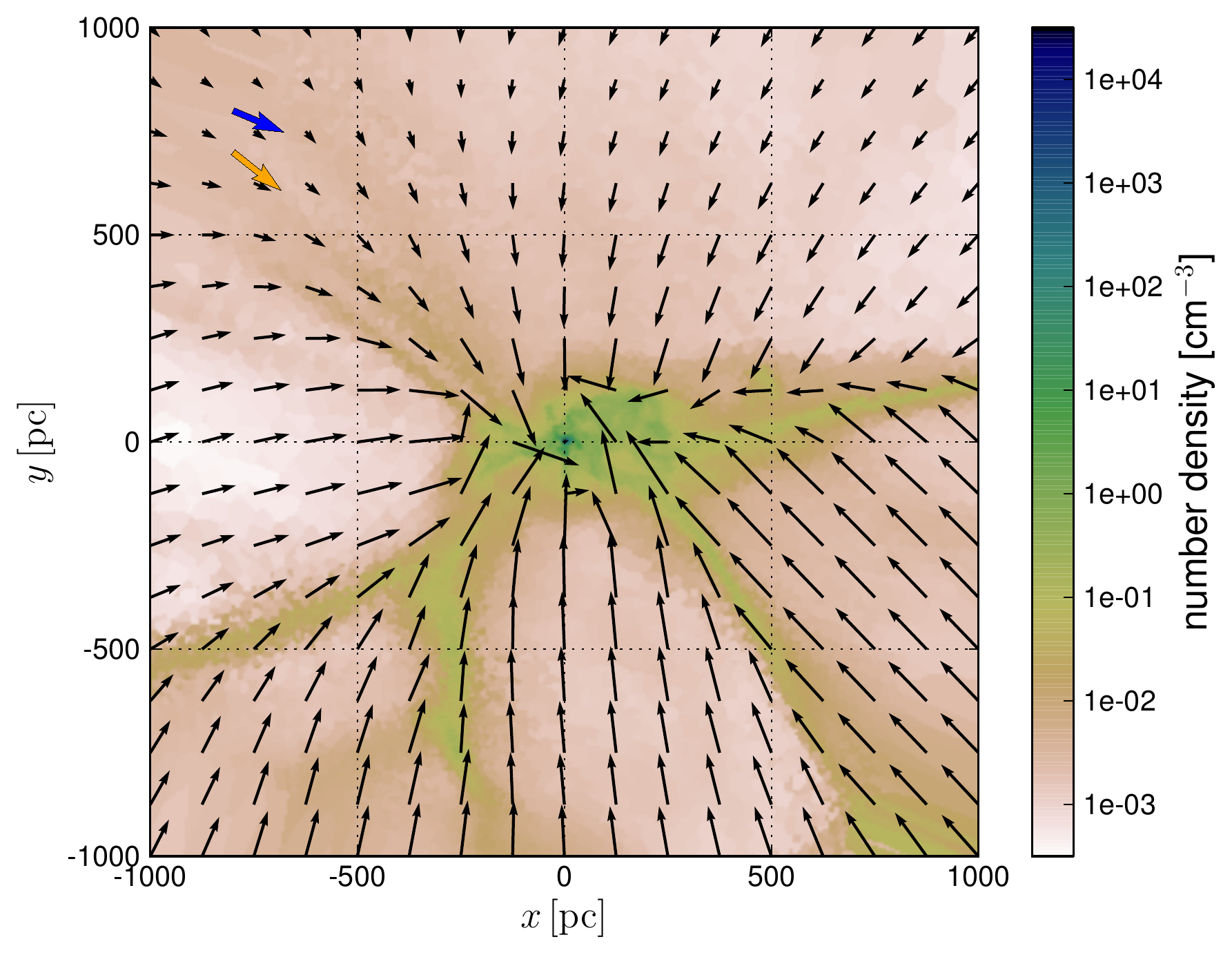}
\includegraphics[width=0.99\columnwidth]{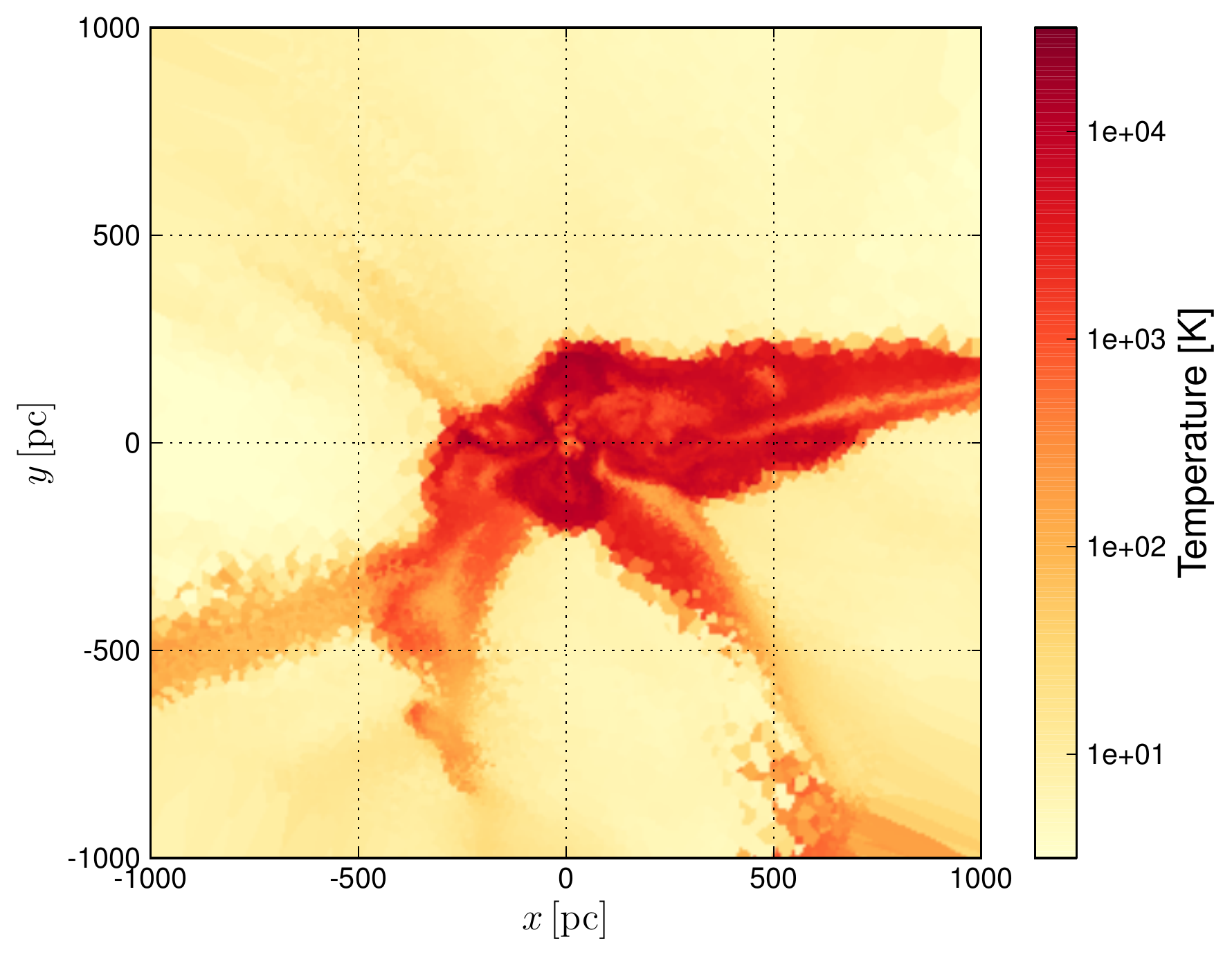}
\caption{Number density (left panels) and temperature (right panels) slices of the collapsed halo 
with M$_\mathrm{halo} = 3 \times 10^7$\,M$_\odot$
in the SB simulation at redshift $z = 14$. 
In the upper two rows, we show the central region in face-on and edge-on projections.
In the bottom row, we show the larger-scale structure surrounding the halo.
The distances shown are in proper units.
The black or white arrows in the density slices show the 
velocity field of the closest cell projected into the plane.  (Note that the velocities shown are computed in the centre of mass rest-frame.) 
In blue, we show the streaming velocity at this redshift, scaled up by a factor of 10, 
and in orange the velocity of the centre of mass. 
We can see that the centre of the halo is dense and cold with temperatures around 
a few hundred K. This is the place where Pop III star formation will eventually take place. 
On larger scales, the gas flows that feed the halo are visible.}
\label{fig:pic}
\end{figure*}

In simulation SB, we form only a single collapsed halo by the end of the simulation. This object has a mass of $3 \times 10^{7} \: {\rm M_{\odot}}$ and a collapse redshift $z_{\rm coll} = 14$. In run LB, on the other hand, the larger box allows us to form more collapsed haloes. We find 2, 3, 10, 13, 37 and 86 haloes that collapse at redshifts $z=19\, , \,18\, ,\,17\, ,\,16\, ,\,15\,$ and 14, respectively.
In every single case, the halo mass at collapse is larger than the mass of a halo with a virial temperature of 8000~K (indicated by the dashed black line in Figure~\ref{fig:coll}), which is a reasonable proxy for the minimum mass of an ACH. 
Previous work has shown that Pop III stars in regions of the Universe with no streaming velocity 
can form in minihaloes with masses of a few times $10^5$\,\msolar \citep{met01,yahs03,hir15}.
We find no cold dense gas in any halo with a mass less than the atomic colling limit, demonstrating that when the baryonic streaming velocity is very high, the formation of Pop III stars in H$_{2}$-cooled minihaloes is very strongly suppressed. 

\begin{figure*}
\includegraphics[width=1.99\columnwidth]{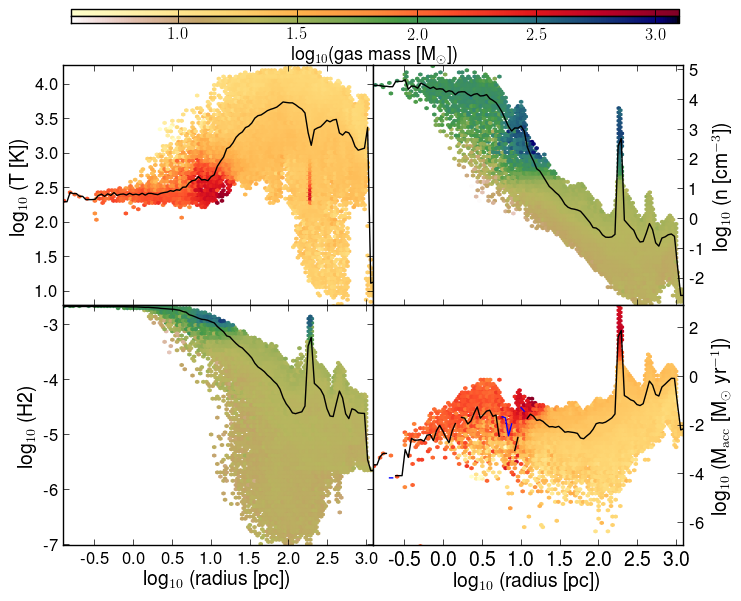}
\caption{Properties of the collapsed halo with M$_\mathrm{halo} = 3 \times 10^7$\,M$_\odot$ 
in the SB simulation at redshift $z = 14$. From top left to  bottom right, we show the temperature, 
gas number density, H$_2$ abundance and mass accretion rate as a function of radius, with the
radius given in proper units. The colored 
regions show two dimensional histograms of the distribution of gas cells, the black lines the mass averaged values.
For the mass accretion rate, we show a blue line for outflowing masses. 
In the central, high density halo region, the temperature drops to a few hundred Kelvin due to 
cooling by H$_2$. In the same region, the H$_2$ abundance reaches values above 10$^{-3}$. Since the accretion rate 
is low ($\approx 10^{-2}$\msolar yr$^{-1}$), we expect Pop III star formation and not the 
formation of a supermassive star that eventually collapses into a DCBH.
}
\label{fig:4panels}
\end{figure*}

Our results demonstrate that high baryonic streaming velocities are a viable mechanism for producing a population of chemically pristine ACHs without the need for a high LW background. A sufficient close pair of such haloes collapsing at very similar redshifts would be an ideal site for the formation of a DCBH \citep{vis14c,regan17}. We have carefully examined whether any of the collapsed haloes in the LB simulation have separations of less than 500~pc in proper units, which is a reasonable upper limit on the required separation, but have found no such close pairs.
However, this is unsurprising, since their space density is expected to be very small. 
In their simulations, \cite{vis14c} find two closely synchronized halo pairs 
in a volume of $\approx 1.7 \times 10^4 \,\mathrm{cMpc}^3$, corresponding to a number density of
$1.2 \times 10^{-4}\,\mathrm{cMpc}^{-3}$, where cMpc denotes comoving Mpc. 
Since the combined volume of simulations SB and LB is much smaller ($209 \,\mathrm{cMpc}^3$), the probability of finding a synchronized pair is only $\sim 2.5$\%.

The fraction of the Universe with streaming velocities of $3\, \sigma_\mathrm{rms}$ or higher
is $\approx 5.9 \times 10^{-6}$. 
If we assume that all high-redshift quasars originate from synchronized halo pairs in regions with $v_{\rm stream} \geq 3\, \sigma_\mathrm{rms}$, then we find a quasar number density  of $0.7 \,\mathrm{cGpc}^{-3}$.  This number is in good agreement with 
the low number density of quasars observed at redshift $z = 6$ of about
$1 \,\mathrm{cGpc}^{-3}$ \citep{fan06b}. 
This statement is based on the assumption that all observed high-redshift quasars can be well observed 
and are not quiescent or hidden in a dust cloud. Thus the real number density of quasars could be higher 
than the value that is observed today.

\subsection{Gas properties within the collapsed haloes}
Next, we take a closer look at the collapsed halo in simulation SB. 
In Figure \ref{fig:pic}, we show slices of the number density and the temperature of the central 
region and the surroundings of our halo at redshift $z=14$.
In the upper left panel, we see the face-on cut-through of the central 20\,pc, where all of the dense, 
collapsed gas  can be found. Here, the densities are high enough for efficient formation of H$_2$, leading to 
strong cooling.  The dense region thus corresponds to a region of low temperature (upper right panel). 
In the edge-on cut-through (middle left and middle right panels), one can see a stream of hot, low-density 
gas flowing onto the halo. 
The larger-scale structure in which the halo is embedded can be seen in the lower two panels. 
It is in the middle of several hot gas streams that feed the minihalo. The central 200\,pc 
region is shock-heated up to $10^4$\,K, except for some high-density streams that 
already cool efficiently.

In Figure \ref{fig:4panels}, we can see the radial profiles of our halo in the SB simulation. 
As can already be seen from the slice plots in Figure \ref{fig:pic}, we see a 
drop in temperature at the centre of the halo, corresponding to a region of high density 
and high H$_{2}$ abundance. Furthermore, in the bottom right hand panel of
Figure \ref{fig:4panels} we see that the mass inflow rate onto the central region is, on average,
well below the threshold value of $\sim 0.1$ \msolar yr$^{-1}$ for forming supermassive stars
\citep[e.g.][]{hos13, schleicher13}. These results confirm the argument put forward by \citet{vis14b} that
high streaming velocities are unable to prevent H$_{2}$ formation in dense gas in ACHs, and demonstrate that the suggestion of \citet{tm14} that streaming velocities alone can create 
the necessary conditions for DCBH formation is incorrect.

\begin{figure}
\includegraphics[width=0.99\columnwidth]{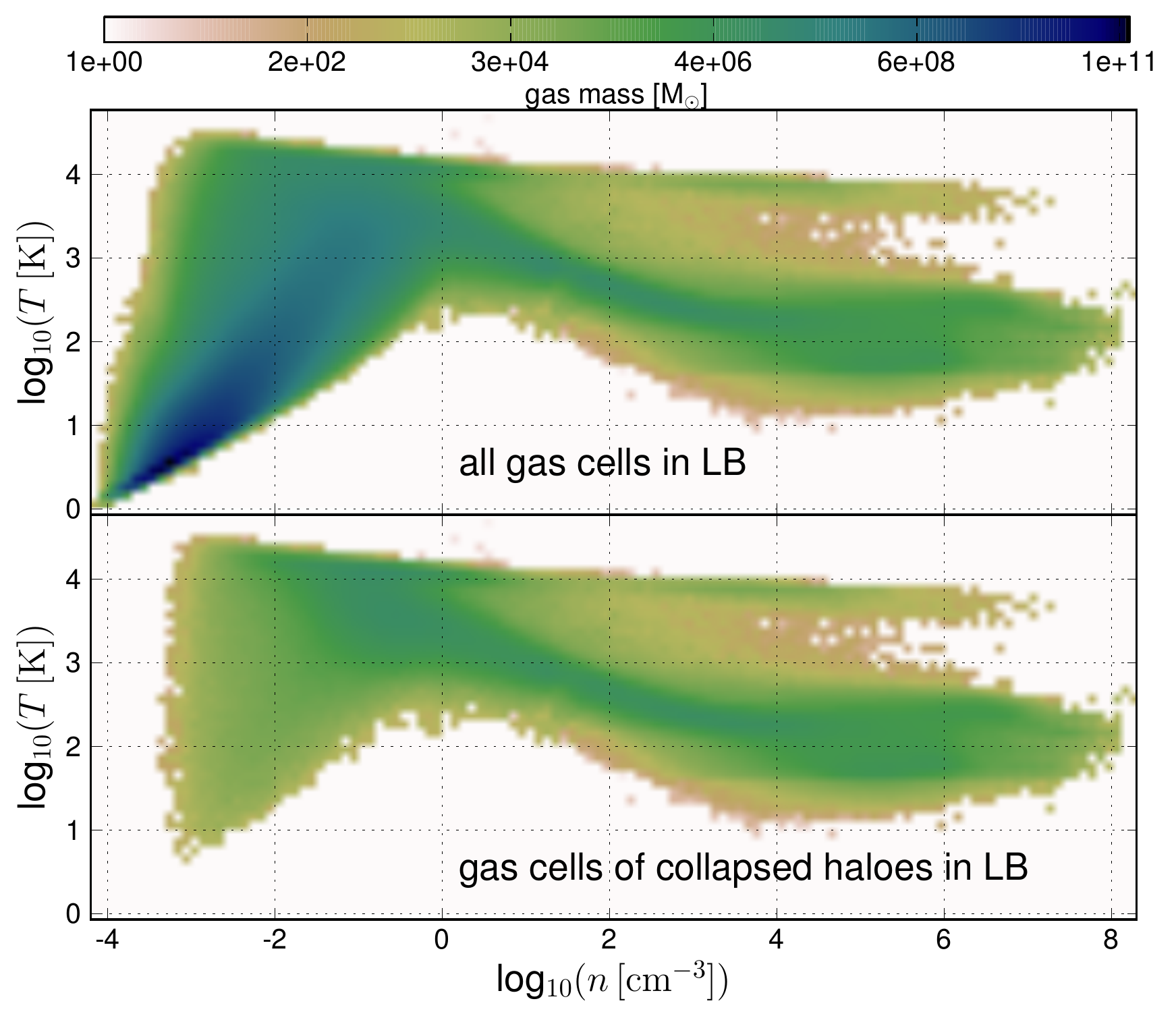}
\caption{Two dimensional histograms of all gas cells in $\log(n)$--$\log(T)$ space at redshift $z=14$. 
The upper panel shows all gas cells in simulation LB, the lower panel all gas 
cells in LB that are associated with the 
collapsed haloes. 
Above number densities of $n \gtrapprox 1$cm$^{-3}$, 
the majority of the gas cells under-go cooling by H$_2$ and 
HD and reach temperatures as low as hundred K. The few high density -- high temperature gas cells can be associated with single shocked gas cells in some haloes. 
We can conclude that all collapsed haloes in simulation LB have cool, dense centres similar to the 
halo in simulation SB.
}
\label{fig:logn-logT}
\end{figure}

In the top panel of Figure \ref{fig:logn-logT}, we show a two-dimensional histogram of all gas cells in simulation LB 
in a $\log(n)$--$\log(T)$ diagram. The region with low number density and temperature corresponds 
to void regions in the simulation, where the gas can cool below the CMB temperature adiabatically. 
In the bottom panel of Figure \ref{fig:logn-logT}, we depict the same histogram, but only for gas cells associated with the 
collapsed haloes at this redshift. Unsurprisingly, there are no cells in the low density, low temperature regime. 

Above gas number densities of $n > 1\, \mathrm{cm}^{-3}$, most of the gas undergoes 
cooling and settles into temperatures of a few hundred K. 
There is, however, some gas that occupies the high density, high temperature region in 
the diagram. 
We carefully check that this hot ($T > 4000$\,K), dense ($n > 10^4\,\mathrm{cm}^{-3}$) gas 
can be traced back to shock heating of individual cells in the haloes \citep[as predicted by][]{io12}.
These cells are typically located at the edge of the high density, cold centre of the halo, where it is intersected by
an inflowing gas stream 
\cite[compare with][who find a similar behaviour in their low LW halo]{latif14b}. 
In the bottom panel of Figure \ref{fig:logn-logT} we see that the majority of the 
dense, collapsed gas in our haloes cools efficiently via H$_2$ and HD emission down to temperatures 
of around a hundred K.  We can therefore safely conclude that all of the collapsed haloes that form in our
simulations have a cold, dense centre,  and are therefore unlikely to collapse to form DCBHs in the
absence of external LW feedback.

\section{Conclusions}

In this paper, we have investigated the role that high baryonic streaming velocities may play in the
formation of DCBHs. We carried out two simulations in which we imposed a constant velocity offset,
$v_{\rm stream} = 3 \sigma_{\rm rms}$, where $\sigma_{\rm rms}$ is the root-mean-squared streaming
velocity. One of these simulations used a volume of side length $1 \, {\rm cMpc}/h$, giving us
very good mass resolution, but producing only a single sample of an ACH. The other simulation used a
larger volume with side length $4 \, {\rm cMpc}/h$, and consequently had worse mass resolution but
yielded a much larger sample of ACHs. In both cases, we find that the effect of the large streaming
velocity is to suppress H$_{2}$ formation (and hence star formation) in minihaloes: the first haloes
in which gas is able to cool and collapse to high densities are all ACHs. However, contrary to a
suggestion by \citet{tm14}, we find no evidence for the suppression of H$_{2}$ cooling in any of the
collapsed ACHs. In every case, H$_{2}$ forms within the centre of the halo, and so in the absence
of additional physics, we would expect all of the haloes to form Pop III stars rather than DCBHs.
In the cases where Pop III star formation is the final result and if the Pop III star(s)
find themselves at the centre of a convergent flow with high in-fall rates then rapid growth may be
possible even when starting from a so-called ``lighter seed''\citep{lupi2016,pezzulli2016}.

Nonetheless, our results demonstrate that high streaming velocities cannot by themselves produce the conditions required for DCBH formation. However, they do produce an ideal environment in which the
synchronised halo model for DCBH formation can operate. This model requires that star formation
be suppressed in a close pair of haloes until both reach masses large enough that they can start to
atomically cool \citep{dijkstra08,vis14c,regan17}. Achieving this with LW feedback is possible,
but requires a LW radiation field strength that is significantly higher than the typical background
value \citep{vis14c}. High streaming velocities provide an alternative means of achieving the same
suppression without the need for a locally elevated LW background, and without the risk that the
haloes will be enriched with metals from nearby minihaloes. In addition, if we estimate the
comoving number density of synchronised halo pairs that
are located in regions with $v_{\rm stream} \geq 3 \sigma_{\rm rms}$, we find a number of order $1 \:
{\rm cGpc}^{-3}$, in good agreement with the observed comoving number density of $z > 6$ quasars.
We conclude that synchronised halo pairs forming in regions of very high baryonic streaming velocity 
are ideal candidates for forming the seeds of the first quasars. 

\section*{Acknowledgments}
The authors would like to acknowledge fruitful discussions with Naoki Yoshida and 
Takashi Hosokawa as well as with Mattis Magg and Mattia Sormani. 
They would like to thank the anonymous referee for her/his valuable comments 
that helped to improve the paper. 
ATPS would like to thank Volker Springel and his team for kindly providing the code 
{\sc arepo} that was used to carry out this simulation. Special thanks goes to 
Christian Arnold, R\"udiger Pakmor, Kevin Schaal, Christine Simpson and Rainer Weinberger 
for their help with the code. 
ATPS, SCOG and RSK acknowledge support
from the European Research Council under the European Community's Seventh Framework
Programme (FP7/2007 - 2013) via the ERC Advanced Grant ``STARLIGHT: Formation of the
First Stars" (project number 339177).
SCOG and RSK also acknowledge support from
the Deutsche Forschungsgemeinschaft via SFB 881 ``The Milky Way System'' (sub-projects
B1, B2 and B8) and SPP 1573 ``Physics of the Interstellar Medium'' (grant number GL 668/2-1).
JR acknowledges the support of the EU Commission through the
Marie Sk\l{}odowska-Curie Grant - ``SMARTSTARS" - grant number 699941. 
The authors gratefully acknowledge the Gauss Centre for Supercomputing e.V. (www.gauss-centre.eu) for funding this project by providing computing time on the GCS Supercomputer SuperMUC at Leibniz Supercomputing Centre (www.lrz.de).
The authors acknowledge support by the state of Baden-W\"urttemberg through bwHPC and the German Research Foundation (DFG) through grant INST 35/1134-1 FUGG.
\bibliographystyle{mn2e}
\setlength{\bibhang}{2.0em}
\setlength\labelwidth{0.0em}
\bibliography{refs}

\begin{thebibliography}{66}
\expandafter\ifx\csname natexlab\endcsname\relax\def\natexlab#1{#1}\fi

\bibitem[{{Agarwal} {et~al}\mbox{.}(2014){Agarwal}, {Dalla Vecchia}, {Johnson},
  {Khochfar}, \& {Paardekooper}}]{agarwal14}
{Agarwal} B., {Dalla Vecchia} C., {Johnson} J.~L., {Khochfar} S.,
  {Paardekooper} J.-P., 2014, \mnras, 443, 648

\bibitem[{{Agarwal} \& {Khochfar}(2015)}]{agar15}
{Agarwal} B., {Khochfar} S., 2015, \mnras, 446, 160

\bibitem[{{Agarwal} {et~al}\mbox{.}(2012){Agarwal}, {Khochfar}, {Johnson},
  {Neistein}, {Dalla Vecchia}, \& {Livio}}]{agarw12}
{Agarwal} B., {Khochfar} S., {Johnson} J.~L., {Neistein} E., {Dalla Vecchia}
  C., {Livio} M., 2012, \mnras, 425, 2854

\bibitem[{{Agarwal} {et~al}\mbox{.}(2017){Agarwal}, {Regan}, {Klessen},
  {Downes}, \& {Zackrisson}}]{agarwal17}
{Agarwal} B., {Regan} J., {Klessen} R.~S., {Downes} T.~P., {Zackrisson} E.,
  2017, MNRAS, submitted; arXiv:1703.08181

\bibitem[{{Agarwal} {et~al}\mbox{.}(2016){Agarwal}, {Smith}, {Glover},
  {Natarajan}, \& {Khochfar}}]{bhaskar15}
{Agarwal} B., {Smith} B., {Glover} S., {Natarajan} P., {Khochfar} S., 2016,
  \mnras, 459, 4209

\bibitem[{{Ahn} {et~al}\mbox{.}(2009){Ahn}, {Shapiro}, {Iliev}, {Mellema}, \&
  {Pen}}]{ahn09}
{Ahn} K., {Shapiro} P.~R., {Iliev} I.~T., {Mellema} G., {Pen} U., 2009, \apj,
  695, 1430

\bibitem[{{Badnell}(2006)}]{b06}
{Badnell} N.~R., 2006, \aap, 447, 389

\bibitem[{{Begelman}, {Volonteri} \& {Rees}(2006){Begelman}, {Volonteri}, \&
  {Rees}}]{begel06}
{Begelman} M.~C., {Volonteri} M., {Rees} M.~J., 2006, \mnras, 370, 289

\bibitem[{{Bromm} \& {Loeb}(2003)}]{bl03}
{Bromm} V., {Loeb} A., 2003, \apj, 596, 34

\bibitem[{{Chandrasekhar}(1964)}]{chandrasekhar64}
{Chandrasekhar} S., 1964, \apj, 140, 417

\bibitem[{{Clark} {et~al}\mbox{.}(2011){Clark}, {Glover}, {Klessen}, \&
  {Bromm}}]{cgkb11}
{Clark} P.~C., {Glover} S.~C.~O., {Klessen} R.~S., {Bromm} V., 2011, \apj, 727,
  110

\bibitem[{{Coppola} {et~al}\mbox{.}(2011){Coppola}, {Longo}, {Capitelli},
  {Palla}, \& {Galli}}]{cop11}
{Coppola} C.~M., {Longo} S., {Capitelli} M., {Palla} F., {Galli} D., 2011,
  \apjs, 193, 7

\bibitem[{{Dijkstra} {et~al}\mbox{.}(2008){Dijkstra}, {Haiman}, {Mesinger}, \&
  {Wyithe}}]{dijkstra08}
{Dijkstra} M., {Haiman} Z., {Mesinger} A., {Wyithe} J.~S.~B., 2008, \mnras,
  391, 1961

\bibitem[{{Eisenstein} \& {Hu}(1998)}]{eh98}
{Eisenstein} D.~J., {Hu} W., 1998, \apj, 496, 605

\bibitem[{{Fan}(2006)}]{fan06b}
{Fan} X., 2006, \nar, 50, 665

\bibitem[{{Fan} {et~al}\mbox{.}(2006){Fan}, {Strauss}, {Richards}, {Hennawi},
  {Becker}, {White}, {Diamond-Stanic}, {Donley}, {Jiang}, {Kim}, {Vestergaard},
  {Young}, {Gunn}, {Lupton}, {Knapp}, {Schneider}, {Brandt}, {Bahcall},
  {Barentine}, {Brinkmann}, {Brewington}, {Fukugita}, {Harvanek}, {Kleinman},
  {Krzesinski}, {Long}, {Neilsen}, {Nitta}, {Snedden}, \& {Voges}}]{fan06}
{Fan} X. {et~al.}, 2006, \aj, 131, 1203

\bibitem[{{Fernandez} {et~al}\mbox{.}(2014){Fernandez}, {Bryan}, {Haiman}, \&
  {Li}}]{fernandez14}
{Fernandez} R., {Bryan} G.~L., {Haiman} Z., {Li} M., 2014, \mnras, 439, 3798

\bibitem[{{Field}, {Somerville} \& {Dressler}(1966){Field}, {Somerville}, \&
  {Dressler}}]{field66}
{Field} G.~B., {Somerville} W.~B., {Dressler} K., 1966, \araa, 4, 207

\bibitem[{{Glover}(2005)}]{glover05}
{Glover} S., 2005, \ssr, 117, 445

\bibitem[{{Glover}(2015)}]{glover15}
{Glover} S.~C.~O., 2015, \mnras, 451, 2082

\bibitem[{{Glover}(2016)}]{g16}
{Glover} S.~C.~O., 2016, arXiv:1610.05679

\bibitem[{{Glover} \& {Abel}(2008)}]{ga08}
{Glover} S.~C.~O., {Abel} T., 2008, \mnras, 388, 1627

\bibitem[{{Glover} \& {Jappsen}(2007)}]{gj07}
{Glover} S.~C.~O., {Jappsen} A.-K., 2007, \apj, 666, 1

\bibitem[{{Greif} {et~al}\mbox{.}(2011){Greif}, {White}, {Klessen}, \&
  {Springel}}]{greif11}
{Greif} T.~H., {White} S.~D.~M., {Klessen} R.~S., {Springel} V., 2011, \apj,
  736, 147

\bibitem[{{Hahn} \& {Abel}(2011)}]{hahn11}
{Hahn} O., {Abel} T., 2011, \mnras, 415, 2101

\bibitem[{{Hartwig} {et~al}\mbox{.}(2015){Hartwig}, {Glover}, {Klessen},
  {Latif}, \& {Volonteri}}]{hartwig15a}
{Hartwig} T., {Glover} S.~C.~O., {Klessen} R.~S., {Latif} M.~A., {Volonteri}
  M., 2015, \mnras, 452, 1233

\bibitem[{{Hirano} {et~al}\mbox{.}(2015){Hirano}, {Hosokawa}, {Yoshida},
  {Omukai}, \& {Yorke}}]{hir15}
{Hirano} S., {Hosokawa} T., {Yoshida} N., {Omukai} K., {Yorke} H.~W., 2015,
  \mnras, 448, 568

\bibitem[{{Hirata} \& {Padmanabhan}(2006)}]{hp06}
{Hirata} C.~M., {Padmanabhan} N., 2006, \mnras, 372, 1175

\bibitem[{{Hosokawa} {et~al}\mbox{.}(2013){Hosokawa}, {Yorke}, {Inayoshi},
  {Omukai}, \& {Yoshida}}]{hos13}
{Hosokawa} T., {Yorke} H.~W., {Inayoshi} K., {Omukai} K., {Yoshida} N., 2013,
  \apj, 778, 178

\bibitem[{{Inayoshi} \& {Omukai}(2012)}]{io12}
{Inayoshi} K., {Omukai} K., 2012, \mnras, 422, 2539

\bibitem[{{Inayoshi} \& {Tanaka}(2015)}]{it15}
{Inayoshi} K., {Tanaka} T.~L., 2015, \mnras, 450, 4350

\bibitem[{{Inayoshi}, {Visbal} \& {Kashiyama}(2015){Inayoshi}, {Visbal}, \&
  {Kashiyama}}]{inayoshi15}
{Inayoshi} K., {Visbal} E., {Kashiyama} K., 2015, \mnras, 453, 1692

\bibitem[{{Latif} {et~al}\mbox{.}(2014b){Latif}, {Bovino}, {Van Borm},
  {Grassi}, {Schleicher}, \& {Spaans}}]{latif14b}
{Latif} M.~A., {Bovino} S., {Van Borm} C., {Grassi} T., {Schleicher} D.~R.~G.,
  {Spaans} M., 2014b, \mnras, 443, 1979

\bibitem[{{Latif}, {Niemeyer} \& {Schleicher}(2014){Latif}, {Niemeyer}, \&
  {Schleicher}}]{lns14}
{Latif} M.~A., {Niemeyer} J.~C., {Schleicher} D.~R.~G., 2014, \mnras, 440, 2969

\bibitem[{{Latif} {et~al}\mbox{.}(2014a){Latif}, {Schleicher}, {Bovino},
  {Grassi}, \& {Spaans}}]{latif14a}
{Latif} M.~A., {Schleicher} D.~R.~G., {Bovino} S., {Grassi} T., {Spaans} M.,
  2014a, \apj, 792, 78

\bibitem[{{Loeb} \& {Rasio}(1994)}]{loeb94}
{Loeb} A., {Rasio} F.~A., 1994, \apj, 432, 52

\bibitem[{{Lupi} {et~al}\mbox{.}(2016){Lupi}, {Haardt}, {Dotti}, {Fiacconi},
  {Mayer}, \& {Madau}}]{lupi2016}
{Lupi} A., {Haardt} F., {Dotti} M., {Fiacconi} D., {Mayer} L., {Madau} P.,
  2016, \mnras, 456, 2993

\bibitem[{{Machacek}, {Bryan} \& {Abel}(2001){Machacek}, {Bryan}, \&
  {Abel}}]{met01}
{Machacek} M.~E., {Bryan} G.~L., {Abel} T., 2001, \apj, 548, 509

\bibitem[{{Mocz} {et~al}\mbox{.}(2015){Mocz}, {Vogelsberger}, {Pakmor},
  {Genel}, {Springel}, \& {Hernquist}}]{mocz15}
{Mocz} P., {Vogelsberger} M., {Pakmor} R., {Genel} S., {Springel} V.,
  {Hernquist} L., 2015, \mnras, 452, 3853

\bibitem[{{Mortlock} {et~al}\mbox{.}(2011){Mortlock}, {Warren}, {Venemans},
  {Patel}, {Hewett}, {McMahon}, {Simpson}, {Theuns}, {Gonz{\'a}les-Solares},
  {Adamson}, {Dye}, {Hambly}, {Hirst}, {Irwin}, {Kuiper}, {Lawrence}, \&
  {R{\"o}ttgering}}]{mort11}
{Mortlock} D.~J. {et~al.}, 2011, \nat, 474, 616

\bibitem[{{Naoz}, {Yoshida} \& {Gnedin}(2013){Naoz}, {Yoshida}, \&
  {Gnedin}}]{naoz13}
{Naoz} S., {Yoshida} N., {Gnedin} N.~Y., 2013, \apj, 763, 27

\bibitem[{{Oh} \& {Haiman}(2002)}]{oh02}
{Oh} S.~P., {Haiman} Z., 2002, \apj, 569, 558

\bibitem[{{Pakmor} {et~al}\mbox{.}(2016){Pakmor}, {Springel}, {Bauer}, {Mocz},
  {Munoz}, {Ohlmann}, {Schaal}, \& {Zhu}}]{pakmor16}
{Pakmor} R., {Springel} V., {Bauer} A., {Mocz} P., {Munoz} D.~J., {Ohlmann}
  S.~T., {Schaal} K., {Zhu} C., 2016, \mnras, 455, 1134

\bibitem[{{Pezzulli}, {Valiante} \& {Schneider}(2016){Pezzulli}, {Valiante}, \&
  {Schneider}}]{pezzulli2016}
{Pezzulli} E., {Valiante} R., {Schneider} R., 2016, \mnras, 458, 3047

\bibitem[{{Planck Collaboration} {et~al}\mbox{.}(2016){Planck Collaboration},
  {Ade}, {Aghanim}, {Arnaud}, {Ashdown}, {Aumont}, {Baccigalupi}, {Banday},
  {Barreiro}, {Bartlett}, \& et~al.}]{Planck15}
{Planck Collaboration} {et~al.}, 2016, \aap, 594, A13

\bibitem[{{Regan} \& {Haehnelt}(2009{\natexlab{a}})}]{rh09}
{Regan} J.~A., {Haehnelt} M.~G., 2009{\natexlab{a}}, \mnras, 396, 343

\bibitem[{{Regan} \& {Haehnelt}(2009{\natexlab{b}})}]{rh09a}
{Regan} J.~A., {Haehnelt} M.~G., 2009{\natexlab{b}}, \mnras, 393, 858

\bibitem[{{Regan}, {Johansson} \& {Wise}(2014){Regan}, {Johansson}, \&
  {Wise}}]{regan14b}
{Regan} J.~A., {Johansson} P.~H., {Wise} J.~H., 2014, \apj, 795, 137

\bibitem[{{Regan}, {Johansson} \& {Wise}(2016{\natexlab{a}}){Regan},
  {Johansson}, \& {Wise}}]{regan16a}
{Regan} J.~A., {Johansson} P.~H., {Wise} J.~H., 2016{\natexlab{a}}, \mnras

\bibitem[{{Regan}, {Johansson} \& {Wise}(2016{\natexlab{b}}){Regan},
  {Johansson}, \& {Wise}}]{regan16b}
{Regan} J.~A., {Johansson} P.~H., {Wise} J.~H., 2016{\natexlab{b}}, \mnras,
  461, 111

\bibitem[{{Regan} {et~al}\mbox{.}(2017){Regan}, {Visbal}, {Wise}, {Haiman},
  {Johansson}, \& {Bryan}}]{regan17}
{Regan} J.~A., {Visbal} E., {Wise} J.~H., {Haiman} Z., {Johansson} P.~H.,
  {Bryan} G.~L., 2017, Nature Astronomy, 1, 0075

\bibitem[{{Sasaki} {et~al}\mbox{.}(2014){Sasaki}, {Clark}, {Springel},
  {Klessen}, \& {Glover}}]{mei14}
{Sasaki} M., {Clark} P.~C., {Springel} V., {Klessen} R.~S., {Glover} S.~C.~O.,
  2014, \mnras, 442, 1942

\bibitem[{{Schauer} {et~al}\mbox{.}(2017){Schauer}, {Agarwal}, {Glover},
  {Klessen}, {Latif}, {Mas-Ribas}, {Rydberg}, {Whalen}, \&
  {Zackrisson}}]{anna17}
{Schauer} A.~T.~P. {et~al.}, 2017, \mnras, 467, 2288

\bibitem[{{Schauer} {et~al}\mbox{.}(2015){Schauer}, {Whalen}, {Glover}, \&
  {Klessen}}]{anna15}
{Schauer} A.~T.~P., {Whalen} D.~J., {Glover} S.~C.~O., {Klessen} R.~S., 2015,
  \mnras, 454, 2441

\bibitem[{{Schleicher} {et~al}\mbox{.}(2013){Schleicher}, {Palla}, {Ferrara},
  {Galli}, \& {Latif}}]{schleicher13}
{Schleicher} D.~R.~G., {Palla} F., {Ferrara} A., {Galli} D., {Latif} M., 2013,
  \aap, 558, A59

\bibitem[{{Shang}, {Bryan} \& {Haiman}(2010){Shang}, {Bryan}, \&
  {Haiman}}]{sbh10}
{Shang} C., {Bryan} G.~L., {Haiman} Z., 2010, \mnras, 402, 1249

\bibitem[{{Springel}(2010)}]{arepo}
{Springel} V., 2010, \mnras, 401, 791

\bibitem[{{Stacy}, {Bromm} \& {Loeb}(2011){Stacy}, {Bromm}, \&
  {Loeb}}]{stacy11a}
{Stacy} A., {Bromm} V., {Loeb} A., 2011, \apjl, 730, L1

\bibitem[{{Stecher} \& {Williams}(1967)}]{stecher67}
{Stecher} T.~P., {Williams} D.~A., 1967, \apjl, 149, L29

\bibitem[{{Sugimura}, {Omukai} \& {Inoue}(2014){Sugimura}, {Omukai}, \&
  {Inoue}}]{sug14}
{Sugimura} K., {Omukai} K., {Inoue} A.~K., 2014, \mnras, 445, 544

\bibitem[{{Tanaka} \& {Li}(2014)}]{tm14}
{Tanaka} T.~L., {Li} M., 2014, \mnras, 439, 1092

\bibitem[{{Tseliakhovich} \& {Hirata}(2010)}]{th10}
{Tseliakhovich} D., {Hirata} C., 2010, \prd, 82, 083520

\bibitem[{{Visbal}, {Haiman} \& {Bryan}(2014{\natexlab{a}}){Visbal}, {Haiman},
  \& {Bryan}}]{vis14b}
{Visbal} E., {Haiman} Z., {Bryan} G.~L., 2014{\natexlab{a}}, \mnras, 442, L100

\bibitem[{{Visbal}, {Haiman} \& {Bryan}(2014{\natexlab{b}}){Visbal}, {Haiman},
  \& {Bryan}}]{vis14c}
{Visbal} E., {Haiman} Z., {Bryan} G.~L., 2014{\natexlab{b}}, \mnras, 445, 1056

\bibitem[{{Wu} {et~al}\mbox{.}(2015){Wu}, {Wang}, {Fan}, {Yi}, {Zuo}, {Bian},
  {Jiang}, {McGreer}, {Wang}, {Yang}, {Yang}, {Thompson}, \& {Beletsky}}]{wu15}
{Wu} X.-B. {et~al.}, 2015, \nat, 518, 512

\bibitem[{{Yoshida} {et~al}\mbox{.}(2003){Yoshida}, {Abel}, {Hernquist}, \&
  {Sugiyama}}]{yahs03}
{Yoshida} N., {Abel} T., {Hernquist} L., {Sugiyama} N., 2003, \apj, 592, 645

\end{thebibliography}

\label{lastpage}

\end{document}